# Evolving Design Rules for the Inverse Granular Packing Problem

Marc Z. Miskin*, Heinrich M. Jaeger

*James Franck Institute and Department of Physics, The University of Chicago, Chicago, IL 60637*

If a collection of identical particles is poured into a container, different shapes will fill to different densities. But what is the shape that fills a container as close as possible to a pre-specified, desired density? We demonstrate a solution to this inverse-packing problem by framing it in the context of artificial evolution. By representing shapes as bonded spheres, we show how shapes may be mutated, simulated, and selected to produce particularly dense or loose packing aggregates, both with and without friction. Moreover, we show how motifs emerge linking these shapes together. The result is a set of design rules that function as an effective solution to the inverse packing problem for given packing procedures and boundary conditions. Finally, we show that these results are verified by experiments on 3D-printed prototypes used to make packings in the real world.

*e-mail: mmiskin@uchicago.edu

It is easy to appreciate the fact that granular aggregates can have different physical properties if the particle shapes are different. For example, it is known that the mechanical strength and stiffness of a granular packing are both influenced by particle shape [1-3]. This follows since particle geometry directly determines both how grains arrange themselves when packed and how stresses propagate throughout the aggregate [3-5]. Yet at present, there is no straightforward way to exploit this relationship and find shapes that produce aggregates with specific, desired stiffness's and/or strengths.

The same problem manifests itself when examining the transport properties of granular packings. Here, microscopic features like the structures of interstices and the number of contacts per particle exert a strong influence on macroscopic properties like fluid permeability[6,7] and thermal/electrical conductivities[7,8]. Again, it is clear that these bulk properties are intimately connected with particle shape, while it remains unclear how to leverage this knowledge to select shapes that tailor aggregates to have preferred behaviors.

The heart of this issue can be put into stark contrast with the two, deceptively similar, questions. First, "Are there particle shapes that pack randomly to different densities?" Clearly the answer is yes [9-14]. Yet change the question to "which particle shapes pack randomly to a density of X?" and a general answer is unknown.

In short, there is an apparent disparity between the number of relationships linking particle shape to aggregate behaviors, and the capacity to transform these same relationships into design rules. In light of this, answering a question like the later one above is tantamount to inventing a method that generically solves the "inverse problem" of associating shape to response. For problems like this, an aggregate's behavior is pre-specified and the goal is to find a particle geometry that produces this behavior.

For a given granular inverse problem, we believe a true solution should come in the form of "design rules." By this, we mean a set of instructions that can be used in and of themselves to easily associate particle shapes to any aggregate behavior that falls within an achievable range. We contrast this with seeking a single geometry that solves a single inverse problem. We seek rules that can be followed to generate solutions to all the inverse problems that can be posed within a specific class.

Here, we introduce a new method that discovers such design rules. This method is built on a new framework that combines artificial evolution, computer simulation, and 3D printing to explore the role of shape in granular matter [15]. In this framework, we generate different shapes by rigidly bonding spheres of different diameters into molecule-like compound particles. Given a "granular molecule", we use discrete element method (DEM) simulations to calculate the properties of its resulting aggregate, and we use these calculated properties to categorize and compare shapes. This allows us to determine how close a given "molecule" comes to achieving a user-specified goal. By selecting and mutating the shapes that come the closest to achieving said goal, we can efficiently and automatically explore the role of particle shape.

In prior work [15] we used this method to identify particles that produce minimum and maximum responses (e.g. particles that produce the stiffest/softest packings). Here we extend this process to build design rules by generating a path of shapes that leads through response space from the minimum response to the maximum response. This path effectively operates as a "design rule solution," at least to the extent that the evolutionary process produces a close approximation to the range of possible behaviors.

The power to achieve this goal comes from our shape representation. Bonded spheres with variable sphere radii are a design space in which any one shape can be morphed into another. By extension, any response can be morphed from one to another, provided the connection between shape and response is continuous.

This process is illustrated schematically in figure 1. Suppose, for example, that a cubic granular molecule needed to be morphed into a sphere. A direct and convenient way to accomplish this is by sending the radius of each sphere in the cube to zero, one at a time. The advantage of this procedure is that a single parameter, i.e. the sum of all sphere radii in the molecule, specifies each intermediate shape uniquely. By making a plot comparing this "shape index" to an aggregate property, designated in figure 1 by Γ, a graphical map emerges that links any response in its range to at least one shape. Because the shape index specifies one and only one geometry, this map is a design rule that functions as an independent inverse problem solution.

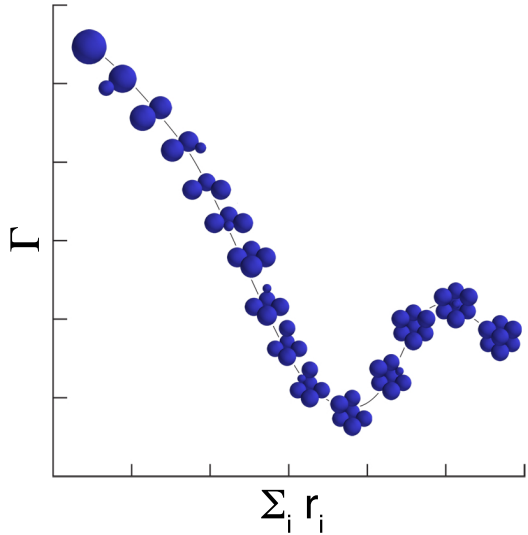

**Figure 1**: A schematic illustrating how a series of deformations in shape can lead to a path of different aggregate responses Γ. In this case, a cube-shaped granular molecule is being transformed into a sphere by systematically sending the constituent sphere radii, $r_i$, to zero. Note that each shape in the path is uniquely specified by the sum of all radii in the granular molecule. Thus, if desired, this graph could be used to solve an inverse problem for the aggregate property Γ: simply draw a horizontal line from the y-axis to the intercept with the response space graph to locate the appropriate shape.

With this in mind, the best design rule is the one that extends the range of responses to be as large as possible. This rephrases the inverse problem as an optimization problem that can be solved via artificial evolution: find the shapes that produce the minimum and maximum responses. The design rule then emerges from considering the path, as in figure 1, that continuously transforms these two extreme shapes into one another.

As a concrete example, here we show how to find particle shapes that pack randomly to a density of X. Since packing fraction is known to depend on both shape and on the role of friction, we consider both packings that are poured in place and packings that are tapped until reaching an asymptotic density. In simulations, we approximate these loading protocols by pouring particles into a box under gravity with frictional and frictionless contact force models (see appendix for details on experiments and simulations).

As the first step to answering this "granular inverse packing problem", we used our evolution framework to identify shapes that generate particularly dense and particularly loose packings when poured under gravity. We allowed our optimizer to make granular molecules from a maximum of 10 constituent spheres of variable radii. Figure 2 shows the median fitness for the simulated populations, under our two loading procedures at each generation. After roughly 50 generations, our algorithm converged when the simulations included friction (Fig 2a) and roughly 100 generations when the simulations were frictionless (Fig 2b).

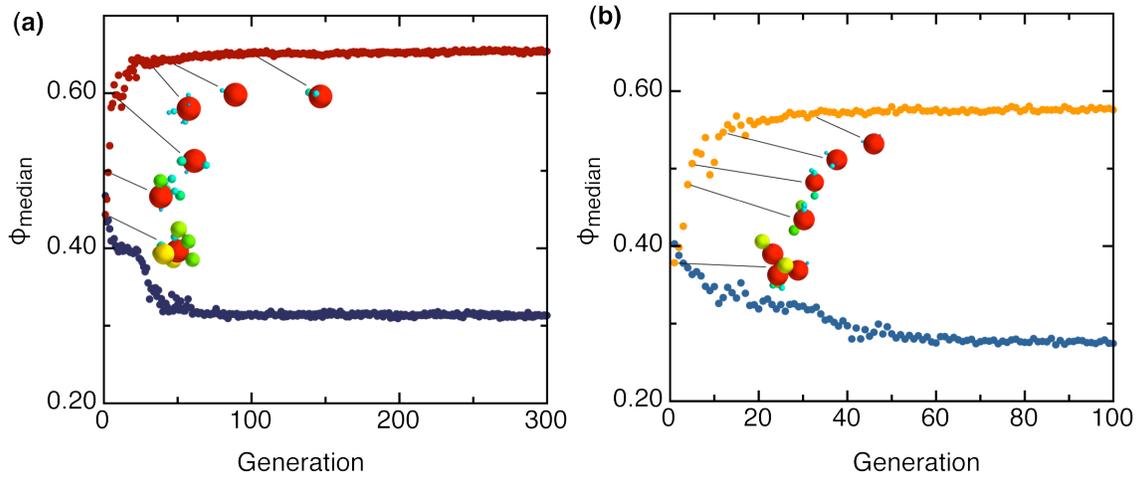

**Figure 2:** Evolving the densest and loosest packing shapes. Starting from random configurations, our evolutionary algorithm adapted particles to pack densely and loosely when poured under gravity without friction (a) and with friction (b). The median packing densities are plotted against generation number. For comparison, we also include call outs to some of the shapes that generated these median values. Note how our evolution process essentially filters out complicated, but suboptimal shapes to discover surprisingly simple shapes as extreme packers.

After starting from the random, un-adapted configurations (Fig. 2), our algorithm identified three distinct geometries (Fig. 3): a sphere, a trimer, and 10mer rod. The trimer shape is symmetric, made from two secondary spheres with radii 0.3 the size of the main sphere, and a bond angle of approximately 70°. These shapes have been adapted to pack densely with friction, densely without friction and loosely, respectively.

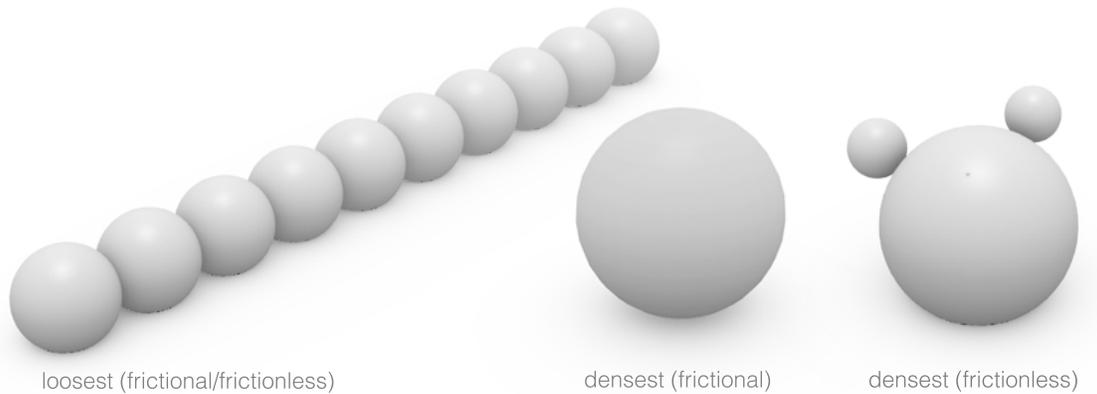

loosest (frictional/frictionless)   densest (frictional)   densest (frictionless)

**Figure 3:** Three extreme packing shapes. The optimizer converged to these three shapes as the particle shapes that generated the most extreme packing densities, given our loading protocol and material parameters. While the rod and trimer exhibit several symmetries, so as to appear relatively simple, note that these are emergent features. They were not programmed as design goals, but rather came about through generations of random mutation and selection.

To verify the results of the optimizer, we 3D-printed particles and measured their packing fractions under the appropriate loading protocols (see appendix). For rods, our simulation predicted packing fractions of 0.27±0.009 and 0.31±0.002 with and without friction. In experiments, we measured packing fractions of 0.26±0.03 and 0.31±0.01 for the as poured packing and the tapped packing, respectively. For spheres, our frictional simulations predicted a packing fraction of 0.578±0.002. Using the "as poured" experimental procedure we measured a volume fraction of 0.59+-0.02, again within error. Finally, we measured a density of 0.67±0.03 for tapped packings of our trimer particle, within error of the simulation prediction, 0.66±0.003.

For the particular optimization goal of finding shapes that minimize or maximize packing density when poured under gravity, the evolution process reduced much of the shape complexity that would have been possible using all 10 spheres. Perhaps surprisingly, all three shapes are relatively simple granular molecules.

Others have identified two of the shapes, rods and spheres, as extreme packing shapes. When considering loose packings prepared by fluidization, Baker and Kudrolli [9] demonstrated that spheres pack denser than any of the Platonic solids. For rods, it has been reported that the volume fraction for an aggregate can be made arbitrarily low by increasing the number of monomers [13, 16]. The fact that the optimizer used all 10 spheres in constructing the rods suggests that here too the packing fraction could have been made lower had we allowed the optimizer more constituent elements.

Perhaps the most surprising shape identified was the trimer particle. We anticipated cubic or tetrahedral granular molecules to be the densest packing shapes without friction. Yet by directly simulating these shapes with our loading procedure and material parameters, we found them to be suboptimal: cubes and tetrahedra made from equal sized spheres packed to 0.465 and 0.641, respectively.

The optimizer's choice became clear after we examined the void space left in a packing of frictionless spheres. Specifically, we studied points that were equidistant to each particle in a set of nearest neighbors (see sketch in figure 4a). If a sphere were placed at one of these points and expanded until just touching all the neighbors, it would be as large as possible without requiring the packing to restructure, making each of these sites local maximum for the size of a spherical void. Therefore, designing granular molecules that fill these voids can potentially be a good strategy to maximize packing density, provided the particles are free to explore different local configurations.

To test this hypothesis we took a packing of spheres of radius $a$ and calculated all such local-maximum void sizes and simulated appending spheres to fill them. We skip any void that overlaps with a void that has been previously added. By making a histogram of the appended sphere radii, d, we found that the most probable void side is 0.3 $a$ (figure 4a). This is the same size ratio as in the trimer particle. What's more, we found a surprising structure in the pair correlation function between these voids. Specifically, we found a local maximum in the probability density corresponding to two voids distributed about 1.5 $a$ apart (figure 4b). Assuming a void size of 0.3 $a$ and that these two voids both touch the same sphere, this would correspond a bond angle of 70°. This is the same bond angle as in our trimer. While there is also a second peak at 2.5 $a$, two 0.3 $a$ appendages can at most be spaced 2.3 $a$ apart, making it impossible to exploit this correlation. Evidently, our trimer was designed to exploit a series of correlations lurking in an ordinary, random packing of spheres.

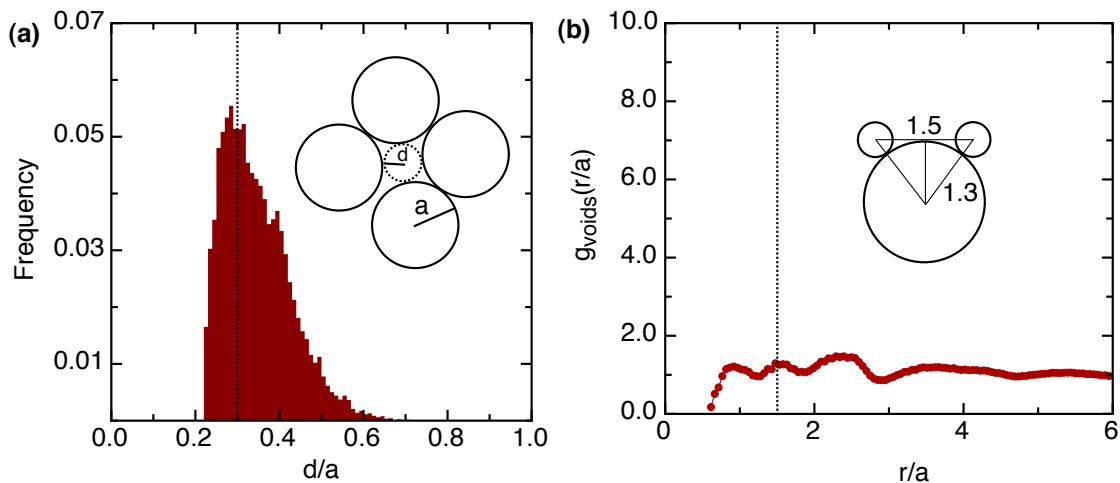

**Figure 4:** Filling the voids of a sphere packing. To rationalize the geometry selected by the optimizer as the densest packing shape, we considered the size (a) and pair correlation (b) of voids in a random packing of spheres of radius a. Voids have a typical size 0.3 *a* and are distributed roughly 1.5 *a* apart. The optimizer picked granular molecules consisting of a central sphere with two nubs that nicely fit into these voids (inset in b).

Combined, the experimental validation and the theoretical rationalization of the shapes found by the optimizer provide a reasonable basis to accept them as good approximations to the true densest and loosest packing shapes. As the next step, we consider how to deform one into the other, and how to use this deformation as a set of design rules.

First, we deform our rod shapes into spheres by sending the sphere radius at the end of the rod to zero, one at a time. We then take the trimer particle and send the secondary spheres to zero radius, again one at a time. These two sets of deformations overlap once the mutated shapes become dimers and are transformed into spheres. Note that, while there are infinitely many ways to perform this deformation, this choice provides a convenient way to index shapes. By systematically sending the furthest spheres to zero, a single number, the sum of all radii in a particle, marks each shape uniquely. For example, if the sum of all radii is 4.2, we know that the shape in question is a linear configuration of four a=1 spheres plus one sphere with a=0.2 sitting at the end.

For each of these mutated shapes, we ran a corresponding DEM simulation, and in figures 5a and 5b we plot the resulting packing fractions against our "shape index," the sum of the radii in the particle. In 5a, the rods form a lower branch spanning all the packing fractions from 0.65 to 0.3 while the trimers yield a second branch, spanning packing fractions from 0.64 to 0.66. For the frictional simulations in figure 5b, rods alone span every packing fraction up to 0.59. Thus, every packing fraction between the minimum and maximum discovered by the optimizer can now be mapped to at least one shape. Furthermore, the shape's geometry can be simply read off the graph, given our deformation procedure.

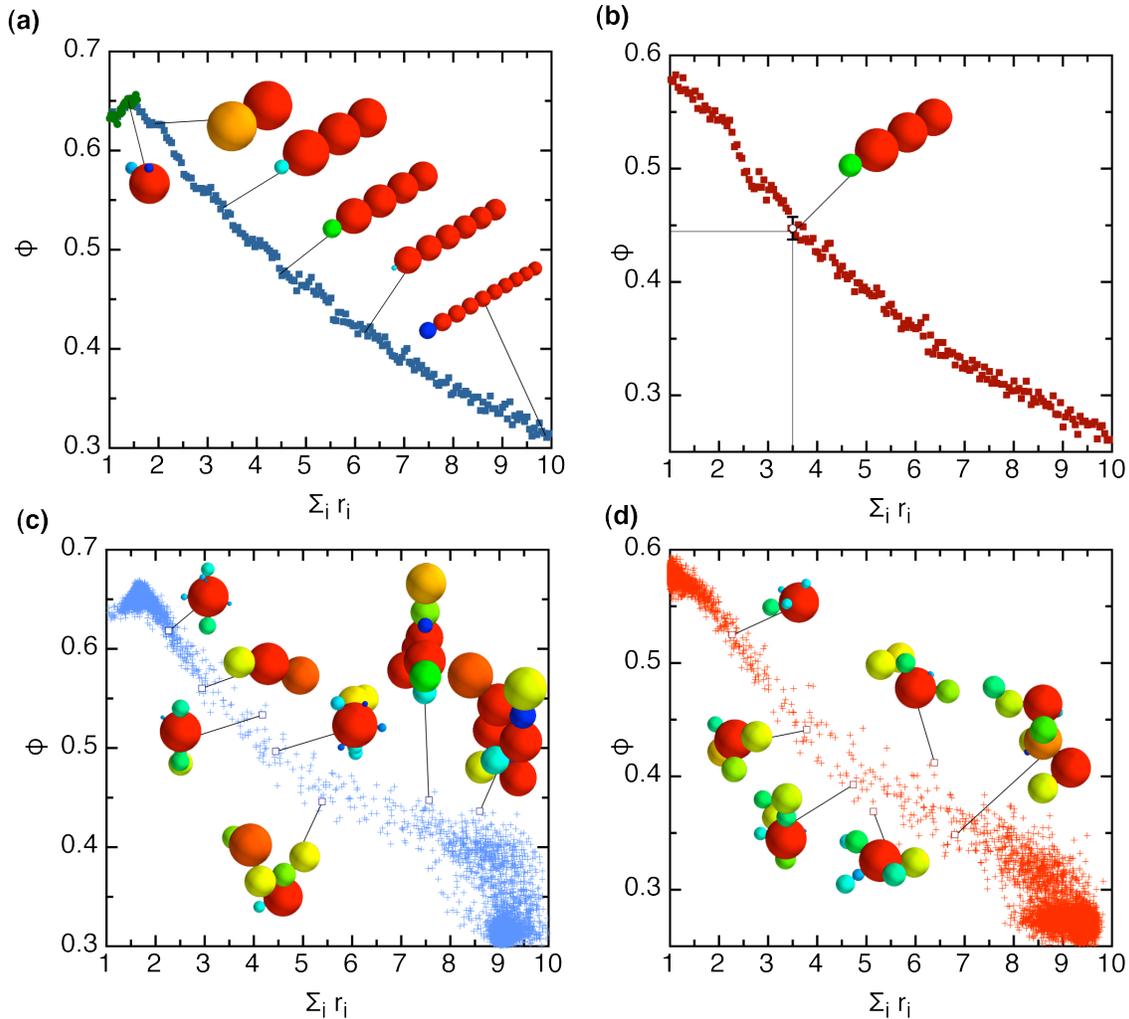

**Figure 5:** Inverting a packing problem. Because the optimizer identified rods, spheres and a trimer as extreme packing shapes, we used these as a basis to build an inverse problem solution. Specifically, we deformed rods (blue in a and red in b) and trimers (green in a), by sending each sphere radius to zero sequentially and simulating the resulting shapes both without (a) and with (b) friction. This procedure generates the plotted paths that can be used to solve inverse problems graphically (b). While this procedure is extremely simple, it is founded on the immense search procedure performed by the optimizer (c,d). For both simulations with and without friction, the optimizer explored a tremendous number of complex and exotic geometries before ultimately converging on the shapes used as a basis for (a) and (b). Any one of these exotic behaviors could, a priori, have been optimal, illustrating the powerful simplifications created by our evolutionary method.

As an illustration of this inverse method in action, we picked an arbitrary packing density, in this case 0.445. Suppose that one wants to find a particle that produces this packing fraction when poured in place. As shown in 5b, the intercept between 0.445 and our response curve occurs for a shape with a shape index of about 3.5. Because the response curve was constructed by sending each sphere to zero sequentially, there is only one shape with this value: a linear particle with three identically sized spheres plus a fourth at the end with one half the radius. We 3D-printed this particle and measured its as poured packing density. We found, as desired, a packing density of 0.45±0.01.

We contrast simple design rules and the simple shapes they lead to against the other shapes considered by the optimizer while perusing the extrema. In figures 5c and 5d we plot all the points it explored in finding the minimum and maximum response shapes with call outs to some of the curious geometries that generated them. It is clear that the optimizer parsed through a tremendous number of exotic and complicated geometries. This large degree of diversity is indicative of how effectively artificial evolution, as a process, filters a broad number of possibilities into an outstanding subset. The optimizer had sufficient evidence to select the shapes in figure 3 as a basis for design rules only after parsing this wide range of potential configurations.

Had we considered packings formed by a protocol different from simple pouring, for example isotropic compression, we may have found different optimal shapes and different maximum/minimum volume fractions. As a case in point in this paper, the inclusion of friction produced different shapes as densest packers. Yet the optimizer operates independently from the simulation procedure at hand, and so these cases would be straightforward generalizations.

Furthermore, our inversion method, based on deforming shapes, can always be used to find a simple graphical map, provided the shape to response mapping is continuous. This is a powerful advantage: it condenses a complicated design problem into a simple set of instructions. Indeed, as shown here, figure 5, can be used by itself as a solution to the inverse granular packing problem, given material and loading procedures similar to ours.

As a first order approximation, the capacity to rationally select shapes that produce desired packing fractions implies the capacity to tune fluid [6] and gas [17] flow through the interstices of a granular bed, thermal and electrical conductivity [8], and a host of other physical properties [7]. However, the relationship between shape and packing fraction explored here serves simply as a proof of principle; any other relationship between particle shape and one or more aggregate properties of the packing could be posed as a granular inverse problem, and solved over some range of values by this method.

The fact that complex, open-ended design problems, like the one examined in this paper, can be reduced to a graph exercise speaks loudly to the power in artificial evolution. In a grand scope, artificial evolution presents a new framework for studying shape, and as this framework is developed and employed, we believe the role of shape in granular systems can be put into focus and put into practice at the same time.

Acknowledgements
We thank Scott Waitukaitis, Juan de Pablo, and Ray Cocco for many insightful discussions. We acknowledge Itasca International Inc. for its generous contribution of the PFC3D software and technical support to this project through the Itasca Education Partnership program. This work was supported by the National Science Foundation through CBET-1334426, with additional support for the DEM simulations from the Nanoscale Science and Engineering Center at the University of Wisconsin through DMR-0832760. M. Z. M. acknowledges support by the US Army Research Office through grant W911NF-12-1-0182 and by a Grainger Graduate Student Fellowship.

**Appendix: Methodology**

Solving a granular inverse problem with our method can be broken down into four steps. First, an experimental procedure is established to quantify the aggregate property at hand, in this case packing density. Next, this experimental procedure is translated into a computer simulation. Any free parameters in the simulation must be calibrated by laboratory measurements. Once the experimental procedure and simulations are established, an evolutionary algorithm is used to identify geometries that lead to extreme behavior. With these results at hand, a path is built through response space and this path functions as the solution to the inverse problem.

*Experimental Protocol:*

Because our design method focuses on particle geometry, it is essential to perform experiments in which shape is the sole variable. This means that material parameters, loading procedures and boundary conditions must be standardized to ensure agreement with simulations.

We fixed all of our material parameters by 3D printing particles in UV cured hard resin (Objet Geometries, VeroWhite). Our particles have a characteristic size of 3mm, and roughly 5,000 copies of each type of compound particle (100mm$^3$ particle volume) were printed at a resolution of 50µm. At this resolution, our printer (Objet Connex 350) guaranteed that, for all practical purposes, particles could be simulated as identical objects made from bonded, perfect spheres.

Next we standardized a definition for volume fraction. We wanted to reduce variability from boundary effects as much as possible, and to do so, we poured samples into different sized and shaped beakers and extrapolated the results to an infinite system size. Our extrapolation method was based on prior tomography results that revealed walls produce a boundary layer of fixed length, l, in which the packing fraction deviates substantially from that of an infinite system[18].

For cylindrical containers, this fixes the dependence of measured packing fraction against system size to functions of the form $\phi_m = \phi_b(1 - \left(1 - \frac{a}{R}l\right)^2) + \phi_c \left(1 - \frac{a}{R}l\right)^2$ where $\phi_b$ is the packing fraction in the boundary layer, $\phi_c$ is the packing fraction in the core region, $a/R$ is the ratio of the largest sphere radius in a granular molecule to the container radius, and l is the boundary layer size, measured in units of a. For spherical containers, the formula essentially the same but with a different exponent: $\phi_m = \phi_b(1 - \left(1 - \frac{a}{R}l\right)^3) + \phi_c \left(1 - \frac{a}{R}l\right)^3$.

We use this function to fit measured packing fractions against system size. To remove dependence on boundary geometry, we performed the analysis by using different shaped containers. Specifically, we made measurements using both spherical and cylindrical vessels. We found the same value for the core density resulted, indicating the extrapolation procedure is appropriate. This best fit value, $\phi_c$ was taken to represent the infinite system size density.

Finally, we needed to account for loading procedure. For any given shape, a range of packing fractions can result depending on whether or not the aggregate has been mechanically agitated or simply poured in place. We elected to consider packings made both as poured and agitated until reaching an approximately steady state. For agitation we used both tapping by hand and mechanical driving. For hand-agitated systems we tapped the container (typically a spherical flask), until no more particles could be added. For mechanically agitated systems we tapped the system (typically a cylindrical container) with approximately 1000 3g (g=9.8m/s$^2$) sinusoidal pulses each spaced by half a second. Measurements of the core region packing fraction with these two methods were consistent with one another, suggesting the packings they generate are good proxies for the asymptotic state.

Below we include a plot of the packing fractions measured in different sized containers, the best-fit lines for our extrapolation procedure, and, for comparison the simulation predictions for the appropriate shape and loading procedure. These data sets were used to generate all the extrapolated packing fractions reported in the main text.

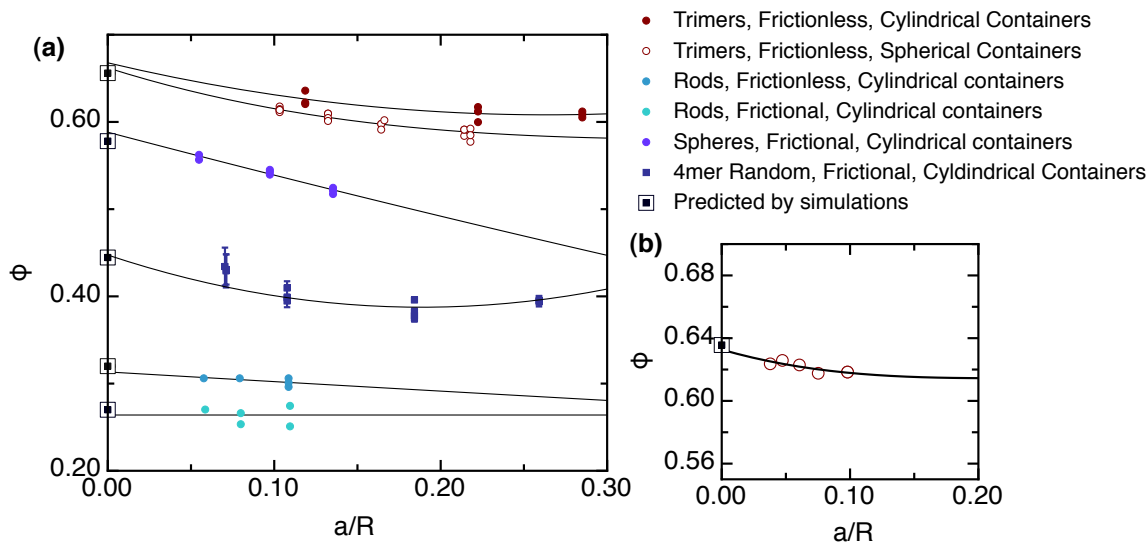

**Fig. S1:** Comparison of packing density from simulations and experiments. The plots show the packing fraction $\phi$ as a function of $a/R$, where $a$ is the size of the largest constituent sphere in a molecule and $R$ is the container radius. (a) Experimental measurements using 3D-printed molecules with lines representing best fits to the model described in the text. Squares along the vertical axis $a/R = 0$ are predictions from the DEM simulations. (b) Measurements on 3mm glass spheres in a spherical container, consolidated by tapping. The extrapolated value $\phi = 0.635$ was used as a validation for the simulation procedure and agrees well with literature values, providing further confirmation of the extrapolation method.

*Simulation protocol*

Our experimental procedure, as defined, required three concepts to be translated into computer simulations. First, the simulations needed to model pouring particles under gravity. Second, the reported value for the packing fraction needed to be independent of system size. Third, we needed simulation parameters that would generate approximations to the as poured packing and the asymptotically tapped packing.

We simulated our packing procedure by settling particles into a box using gravity. The length dimensions in the simulations were set relative to the diameter of a sphere with a volume equal to a single granular molecule. In every simulation, we set 1000 granular molecules into a long, loose column extending 40 diameters from the bottom of the container in which no particles overlap. We chose the height of 40 diameters arbitrarily, although we found no change in packing fraction when we simulated the same procedure with a column of 50 diameters. We then activated gravity causing the particles to settle into a packing. The settling process was stopped when the average distance between particles and the bottom of the box changed by less than $10^{-5}$% and the ratio of the average unbalanced force to the average contact force was less than 1%. We found that lower tolerances did not change the calculated packing fractions.

Once a packing formed, we needed to extract a measurement that represented the infinite system size packing fraction. While, in principle, we could have simulated different sized systems and extrapolated the results, in analogy with our experimental method, we felt this was needlessly costly. Running multiple simulations at different, larger system sizes would significantly increase run time. Furthermore, any single simulation provides us with complete information about particle locations, and thus, gave us a direct way to measure core-region packing fractions. Specifically, we used the particle locations to generate the radical Voroni tessellation for the packing [19]. Then, we made a volume cut by drawing a new, imaginary rectangular box centered on the center of mass. The x, y, and z, dimensions of the box were set equal to the 60$^{th}$ percentile of particle distances measured along each respective axis. We defined the packing fraction by adding up the particle volumes and the Voroni volumes that fall within this box, and dividing the former by the latter.

This simulation method had a number of free model parameters that needed to be appropriately selected to produce good approximations to real packings. Our particles interacted via both normal-force linear springs, and springs that produced tangential forces. The latter, intended to approximate friction, were restrict to never have a magnitude greater than μ times that of the normal springs. Following Cundall and Strack [20], we also included a numerical, global damping to dissipate energy.

When using our simulation protocol to approximate as poured packings, we set our linear spring constant to $2 \times 10^8$ N/m and the frictional tangential spring constant to $4 \times 10^7$ N/m. The frictional slip coefficient, μ, was set to 0.25. These material parameters were set using the procedure detailed in Ref. [15]. Finally, the numerical damping was configured so that, if two particles collide, 20% of the energy would be lost after a loading-unloading cycle. When using our simulation protocol to approximate asymptotically tapped packings, we simply set the tangential spring constant to zero.

While there are other possible contact models, dissipation mechanisms or simulation protocols, we accepted this protocol because it produced reliable agreement with our experimental procedures. For validation, we printed spheres and dimers and measured the infinite system size packing fractions with our extrapolation method. We compared the results to those generated by our simulation procedure. For spheres, as poured, we measured a volume fraction of 0.59±0.02. Our simulation protocol calculated a volume fraction of 0.578±0.002, in good agreement. For spheres tapped into the asymptotic state, we measured a volume fraction of 0.63±0.02 while our frictionless simulations measured a volume fraction of 0.635±0.004. Note both of these values are also in good agreement with literature [21]. For dimers, we measured the as poured volume fraction to be 0.56±0.02 and our frictional simulations calculated a packing fraction of 0.55±0.004.

All the simulations were performed using the PFC3D DEM package from the Itasca Consulting Group, (PFC3D 4.0). The code was executed on a small computing cluster made of 10 identical desktop computers.

*The Evolutionary Algorithm*

With our simulations set to produce accurate approximations to real world packings, we required an external input that would decide what shapes to simulate. Ideally, this input needed to automatically generate new shapes solely on the basis of previous simulations. Furthermore it needed to use this information to discover the particularly novel and interesting shapes that generate extraordinary aggregate properties.

In our methodology, artificial evolution provides this input. Our algorithm of choice for shape exploration is the covariance matrix adaptation evolution strategy (CMA-ES) [22]. In this algorithm, a population undergoing evolution is represented as a multivariate Gaussian distribution. Samples are drawn from this distribution, evaluated for fitness and ranked. This ranked list is used to update the mean and the covariance matrix of the distribution. The mean is updated to maximize the likelihood of successful solutions in the subsequent distribution, whereas the covariance matrix is updated to maximize the likelihood of successful changes to the search distribution.

Repeatedly, the CMA-ES has demonstrated itself as a state of the art optimizer for complex, non-linear, non-separable problems [23]. However, its performance is, as with all optimizers, dependent upon problem representation: it is important that parameters are encoded with equal relevancy, and any boundaries on the search space are implemented as naturally as possible.

We developed a successful problem representation for granular molecules by mapping ordered lists of bearings to particle geometries under what we called blueprint rules [15]. First, a sphere is placed at the origin. Subsequently, each bearing is then taken in turn from the list, and used to draw a ray from the origin along the direction it specifies. Another sphere is then slid along this ray from infinity towards the origin until it touches the object built up in previous steps. This sphere is then fixed in place, and the next bearing is selected. The process iterates until every bearing in the list has been used to place a sphere.

To enrich this framework, we extended these rules to include the sphere radii as variable parameters. This generalization both broadens the range of shapes explored, and continuously encodes the number of sphere elements: since a sphere can always be eliminated by sending its radius to zero, molecules initialized to have, for example, 10 spheres can be transformed into molecules made of 2 spheres, and vice versa. This means that the optimizer can effectively decide how many spheres are needed to produce an optimal shape, given a user-specified maximum. We limited the total number of spheres per granular molecule to the upper bound, *n=10*. This choice was only due to the amount of computational resources that were available.

Since all the variables fall within a fixed range, we imposed boundaries on the CMA-ES search [24]. The radii were forced to range from zero to one while two angles used to specify a bearing were forced to lie between 0 to 360 degrees and 0 to 180 degrees.